\documentclass[aip,apl,reprint,floats,citeautoscript]{revtex4-1}

\usepackage{graphicx}
\usepackage{amsmath}
\usepackage{amssymb}
\usepackage{bm}
\usepackage{times}
\usepackage{colordvi}

\graphicspath{{.}{./EPS/}}

\newcommand{\av}[1]{\langle #1\rangle}

\newcommand*{\vek}[1]{{\ensuremath{\bm{\mathrm{#1}}}}}

\begin{document}

\title{Charge transport by modulating spin-orbit gauge fields for quasi-onedimensional
holes}

\author{T. Kernreiter}
\affiliation{School of Chemical and Physical Sciences and MacDiarmid Institute
for Advanced Materials and Nanotechnology, Victoria University of Wellington,
PO Box 600, Wellington 6140, New Zealand}

\author{M. Governale}
\affiliation{School of Chemical and Physical Sciences and MacDiarmid Institute
for Advanced Materials and Nanotechnology, Victoria University of Wellington,
PO Box 600, Wellington 6140, New Zealand}

\author{A.~R. Hamilton}
\affiliation{School of Physics, University of New South Wales, Sydney, NSW 2052,
Australia}

\author{U. Z\"ulicke}
\affiliation{School of Chemical and Physical Sciences and MacDiarmid Institute
for Advanced Materials and Nanotechnology, Victoria University of Wellington,
PO Box 600, Wellington 6140, New Zealand}

\date{\today}

\begin{abstract}
We present a theoretical study of ac charge transport arising from adiabatic temporal
variation of zero-field spin splitting in a quasi-onedimensional hole system (realized,
e.g., in a quantum wire or point contact). As in conduction-electron systems, part of
the current results from spin-dependent electromotive forces. We find that
the magnitude of this current contribution is two orders of magnitude larger for holes
and exhibits parametric dependences that make it more easily accessible experimentally.
Our results suggest hole structures to be good candidates for realizing devices where
spin currents are pumped by time-varying electric fields.
\end{abstract}


\maketitle 

\begin{figure}[b]
\includegraphics[width=2.7in]{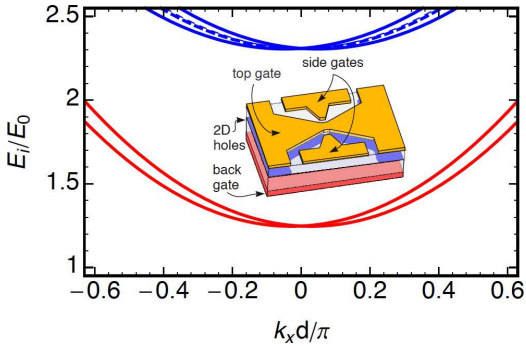}
\caption{\label{fig:setup}
Inset: Setup of a charge-pumping device. Constant side-gate voltages are used
to adjust the point-contact width $w$. Time-varying top- and/or back-gate voltages
modulate the SIA (Rashba) spin-splitting strength (measured in terms of a voltage
$V_z$) in the quantum well of width $d$. Main panel: Dispersion relation of the
lowest quasi-1D subbands in zero magnetic field for aspect ratio $w/d=1$, using
band-structure parameters applicable to GaAs [$\bar{\gamma}\equiv(2\gamma_2
+3\gamma_3)/(5\gamma_1)=0.37$] and $V_z = 0.1 \, V_{\text{R}}$. The voltage
scale $V_{\text{R}}=\gamma_1\pi\hbar^2/(2 m_0 \big|r_{41}^{8v8v}\big|)$ quantifies
the magnitude of spin-orbit effects in a material's band structure.}
\end{figure}
In conventional electric circuits, flow of charges is induced by applied
electric-potential differences. As an alternative particularly suited for miniature
electronic devices, pumping of charge by manipulating intrinsic system
parameters has attracted a lot of attention. Several intriguing proposals of this
kind~\cite{mich:prb:03,chao:prb:03,*avishai:prl:10} are based on the
tunability~\cite{schaep:prb-rc:97,*nitta:prl:97,*grund:prl:00} of spin splittings
induced by structural inversion asymmetry (SIA) in semiconductor
nanostructures~\cite{rolandbook}. Such exotic effects could be studied
experimentally in materials with intrinsically large spin splitting. Generally,
\textit{p}-type structures are favorable~\cite{rolandbook,romain:apl:06,
*oleh:apl:06,*rokh:prl:06,*shay:apl:08,*csontos:apl:10,*gg:nphys:10} because
valence-band states are directly affected by the atomic spin-orbit coupling, while
conduction-band states experience a sizable SIA spin splitting only in narrow-gap
materials. Also, the larger effective mass of holes implies that spin splitting is a
larger fraction of the Fermi energy at given carrier density than in electron
systems.

Here we present a study of charge transport induced by time-dependent SIA
spin splitting in \textit{p}-type (hole) quantum wires or point contacts, taking
into account the special properties of hole states due to their effective spin-3/2
degree of freedom~\cite{YuCardona}. Figure~\ref{fig:setup} illustrates the basic
device setup considered in our work. A quasi-one-dimensional (quasi-1D) hole
system with free propagation direction for holes parallel to the $x$ axis is realized
by engineering the appropriate quantum confinement in the orthogonal in-plane
($y$) direction of a \textit{p\/}-type semiconductor heterostructure that has growth
direction parallel to the $z$ axis. SIA spin splitting is modulated in time by means
of front- and back-gate voltages~\cite{schaep:prb-rc:97,*nitta:prl:97,*grund:prl:00}
that generate an electric field parallel to the heterostructure-growth ($z$) direction.
As for conduction electrons~\cite{mich:prb:03}, the time-dependent SIA spin
splitting acts as a time-dependent and spin-dependent gauge field. The resulting
spin-dependent electromotive forces generate oppositely directed and spin-polarized
partial currents from spin-split subbands. To convert the induced spin current into a
more easily measurable charge current, an in-plane magnetic field is applied
perpendicularly to the quasi-1D system (i.e., in $y$ direction). As we will see below,
the larger magnitude and favorable sample-parameter dependences of this current in
hole systems make it more easily experimentally accessible than in \textit{n}-type
structures.

We use the Luttinger model~\cite{luttham2} for the top-most valence band in
common semiconductor materials. Within the spherical
approximation~\cite{bald:prb:73}, the corresponding effective-mass
Hamiltonian for holes reads (we count hole energies as positive)
\begin{equation}
H_{\text{L}}=\frac{\hbar^2\gamma_{1}}{2 m_{0}}
\left[
\left( 1 + \frac{5}{2}\,\bar\gamma \right) \vek{k}^2 \, \openone  - 2\bar{\gamma}
\left( \vek{k} \cdot \hat{\vek{J}} \right)^2 \right] \quad .
\label{eq:LuttingerHam}
\end{equation}
Here $m_0$ is the electron mass in vacuum, and the parameter $\bar{\gamma}
=(2\gamma_2+3\gamma_3)/(5\gamma_1)$ measures the splitting between
heavy-hole (HH) and light-hole (LH) bulk-valence-band dispersions in terms of
the materials-dependent Luttinger parameters~\cite{vurg:jap:01}. $\vek{k}$
denotes the hole wave vector, and $\hat{\vek{J}}$ is the vector of spin-3/2 
matrices~\cite{rolandbook}. The Zeeman coupling of hole spin to a magnetic field
$\vek{B}$ is described by $H_{\text{Z}}=2 \kappa \mu_{\text{B}} \vek{B}\cdot
\hat{\vek{J}}$, where $\mu_{\rm B}$ denotes the Bohr magneton and $\kappa$ is
the isotropic hole g-factor.\footnote{We neglect the typically very small anisotropic
contribution to Zeeman splitting in the valence band.} SIA gives rise to a coupling
between the holes' orbital motion and their spin. The corresponding (Rashba)
term in the Hamiltonian reads~\cite{rolandbook} $H_{\text{R}} = r_{41}^{8v8v}\,
( \vek{k}\times\vek{\mathcal E} ) \cdot\hat{\vek{J}}$. The band-structure parameter
$r_{41}^{8v8v}=14.62\, e$\AA$^2$ in GaAs (values for other semiconductor
materials can be found in Ref.~\onlinecite{rolandbook}), and $\vek{\mathcal E}$ is 
the effective electric field quantifying SIA.

Following the spirit of previous proposals for quantum pumps of conduction-band
electrons~\cite{mich:prb:03,chao:prb:03}, we consider the holes to be confined in 
a quasi-1D system by a quantum-well confinement in $z$ direction and additional 
in-plane confinement $V(y)$. For simplicity, we assume the confining potential in
both directions to be of hard-wall type, with respective widths $d$ and $w$, and 
project onto quasi-1D bound states by setting $k_z\to\av{k_z}=0$, $k_y\to\av{k_y}
=0$, $k_z^2\to\av{k_z^2} = m^2 \pi^2/d^2$, and $k_y^2\to\av{k_y^2} = n^2 \pi^2/
w^2$, with integer $m$ and $n$. This yields an effective Hamiltonian $H^{(mn)}
(k_x)$ describing HH-LH splitting and mixing within the subspace of a single
quasi-1D bound-state level. As SIA will typically only be sizable in the
quantum-well growth direction, we have $\vek{\mathcal E} = {\mathcal E}_z
\hat{\vek{z}}$. Defining the abbreviation $\hat J_\pm = (\hat J_x \pm i \hat J_y) /
\sqrt{2}$, as well as a natural energy scale $E_0=\pi^2 \hbar^2\gamma_1/(2 m_0
d^2)$, magnetic-field strength $B_0 = \gamma_1\pi^2\hbar/(2 \kappa e d^2)
\approx 18.9\, \text{kT}/(d[\text{nm}])^2$ for GaAs, the SIA voltage scale
$V_{\text{R}}=\gamma_1\pi\hbar^2/(2 m_0 \big|r_{41}^{8v8v}\big|)$ ($\approx 5.72$~V 
for GaAs), and $V_z = {\mathcal E}_z d$, we find  $H^{(mn)}(k_x) = E_0
\big[{\mathcal H}_{\text{qw}}^{(m)} + {\mathcal H}_{\text{pc}}^{(n)} + {\mathcal 
H}_{\text{1D}}(k_x) + {\mathcal H}_{\text{RZ}}(k_x)\big]$, with
\begin{subequations}
\begin{eqnarray}
{\mathcal H}_{\text{qw}}^{(m)} &=& \left[ \openone - 2\bar\gamma \left( \hat
J_z^2 - \frac{5}{4}\,\openone \right)\right] m^2\quad , \\
{\mathcal H}_{\text{pc}}^{(n)} &=& \left[ \openone + \bar\gamma \left( \hat
J_z^2 - \frac{5}{4}\,\openone + \hat J_+^2 + \hat J_-^2 \right) \right] \left(
\frac{n d}{w} \right)^2 , \\
{\mathcal H}_{\text{1D}}(k_x) &=& \left[ \openone + \bar\gamma \left( \hat
J_z^2 - \frac{5}{4}\,\openone - \hat J_+^2 - \hat J_-^2 \right) \right] \left(
\frac{k_x d}{\pi} \right)^2 , \hspace{0.4cm} \\
{\mathcal H}_{\text{RZ}}(k_x) &=& - \frac{V_z}{V_{\text{R}}} \,
\frac{k_x d}{\pi} \, \hat J_y + \frac{\vek{B}}{B_0} \cdot \hat{\vek{J}} \quad .
\end{eqnarray}
\end{subequations}
Straightforward diagonalization of $H^{(mn)}$ yields the spin-split quasi-1D
hole-subband dispersions. See Figure~\ref{fig:setup}.

HH-LH mixing between different orbital bound-state levels is neglected within
our model. Such a simplified approach yields reliable results only for the lowest
spin-split subband~\cite{*[{See, e.g., }] [{, where a similar model for hole point
contacts was discussed.}] uz:pssc:06}. Hence, we limit our study to the case
where only the two subbands with $m=n=1$ are occupied. Thus four
propagation channels exist for holes in the device. Their corresponding Fermi
wave numbers, denoted by $k_{1,2}^{(+)}$ for right-movers and $k_{1,2}^{(-)}$
for left-movers, are straightforwardly found from the subband dispersions.

To model the not necessarily perfect transmission of a hole point contact, we
introduce a $\delta$-barrier potential $V_0\delta(x)$ half-way between the two
leads that make contact to the quasi-1D hole system of length $L$. The
scattering matrix $\mathcal S$ is found by adapting standard methods~\cite{datta}
to the case of four-spinors describing spin-3/2 hole states. 

Application of (front and/or back-gate) voltages can be used to effect a change in 
the SIA spin splitting~\cite{schaep:prb-rc:97,*nitta:prl:97,*grund:prl:00} and should
thus make it possible to impose a variation of $V_z$ with time $\tau$.
Under certain conditions, transport of charge and/or spin will be the result of such
a time dependence~\cite{mich:prb:03,chao:prb:03,*avishai:prl:10}. To produce a
dc charge current in the adiabatic regime, temporal variation of at least
\textit{two\/} parameters is necessary, whereas a time-dependent $V_z$ by itself
only generates ac currents. Studying the latter will provide useful insight into the
physical mechanism for spin-dependent quantum pumping and is also more
easily possible experimentally. Hence we focus here on ac charge transport generated
in situations where only SIA spin splitting is time-dependent.

Using techniques introduced in Ref.~\onlinecite{butt:zpb:94} (see also
Ref.~\onlinecite{brouwer:prb:98}), currents injected in the leads are given
in terms of contributions due to transmission and reflection coefficients. We 
label different blocks of the scattering matrix by $s\in\left\{r,r',t,t'\right\}$ and
define partial currents $i_s$ as\footnote{These partial currents satisfy Onsager-type
relations $i_{t^{(\prime)}}(B_x)=i_{t^{(\prime)}}(-B_x)$, $i_{t}(B_y)=
i_{t^{\prime}}(-B_y)$, $i_{r^{(\prime)}}(B_y)=i_{r^{(\prime)}}(-B_y)$, $i_{r}(B_x)
=i_{r^{\prime}}(-B_x)$, where the last two equations are valid for symmetric
devices with the barrier placed midway between contacts.} 
 \begin{align} 
 \label{eq:is}
i_{s}(\tau)=\frac{e}{2\pi} \, \dot{V}_z(\tau) \sum^2_{i,j=1}\Im m\left[
\frac{\partial s_{ij}}{\partial V_z} {s _{ij}}^*\right] \quad .
\end{align}
Currents entering the left (L) and right (R) leads are then given by $i_{\text{L}} 
(\tau)=i_{t^\prime}(\tau)+i_{r}(\tau)$ and $i_{\text{R}}(\tau)=i_{t}(\tau)+i_{r^{\prime}}
(\tau)$.
We consider the linear combination of left and right currents that is related to net charge
transport between the two leads: $i_{\text{LR}}(\tau)=[i_{\text{L}}(\tau)-i_{\text{R}}(\tau)]/2$. 

For our chosen device geometry, $i_{\text{LR}}$ is entirely due to spin-dependent
electromotive forces. These arise because the wave function of charge carriers travelling
between the leads acquires a dynamical phase that appears in the transmissions:
$t_{jj}\propto\exp(i k^{(+)}_{j} L)$ and $t^{\prime}_{jj}\propto\exp(-i k^{(-)}_{j} L)$. As the
wave vectors $k^{(\pm)}_{j}$ are functions of $V_z$, they are time-dependent when SIA
spin splitting is modulated. Time-dependent phases of transmission coefficients can be
interpreted as electromotive forces that, in the present case, are also subband-dependent.
This is the origin of the spin-orbit gauge-field contribution to  $i_{t^{(\prime)}}$, which
scales linearly with $L$. It turns out that all other contributions to $i_{\text{LR}}$ cancel,
and that the current $i_{\text{LR}}$ is antisymmetric in $B_y$. The antisymmetry of
$i_{\text{LR}}$ w.r.t.\ $B_y$ is the distinctive feature of charge transport generated by
time-dependent spin-orbit gauge fields. This is so because spin-dependent electromotive
forces generate oppositely directed partial currents from the two spin-split subbands
whose imbalance in $B_y\ne 0$ results in a finite transport current $i_{\text{LR}}$.  

\begin{figure}[b]
\includegraphics[width=2.7in]{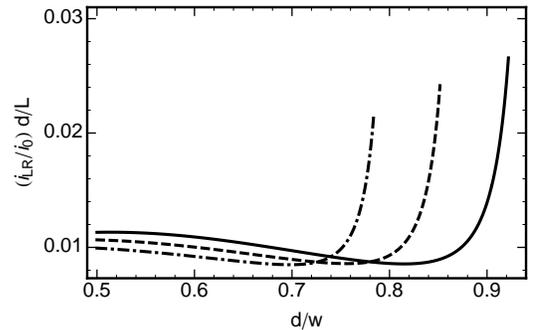}
\caption{\label{fig:obsCurrent}
Charge current between L(eft) and R(ight) leads generated by time-varying
spin-orbit gauge fields, plotted as a function of the aspect ratio $d/w$ for values
of the Fermi energy $E_{\text{F}}= 1.0\, E_0$ (dot-dashed curve),  $1.1\, E_0$
(dashed curve), $1.2\, E_0$ (solid curve). Other parameters are $\bar\gamma
=0.37$ (the value for GaAs holes), $V_0 = 0.1\, E_0 d$, $V_z=0.1\,
V_{\text{R}}$, $B_x=0$, and $B_y = 0.02\, B_0$. $i_0 = e \dot{V}_z / V_{\text{R}}$.
Currents are scaled by the factor $d/L$.}
\end{figure}
\begin{figure}[t]
\includegraphics[width=2.7in]{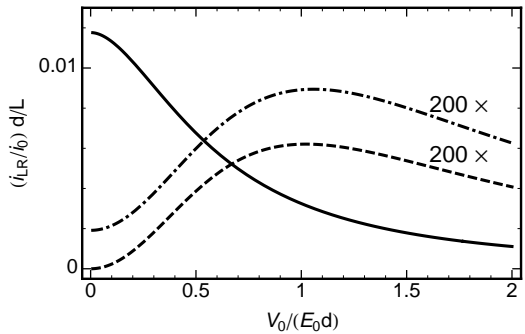}
\caption{\label{fig:IV}
Dependence of  ac charge current on the strength $V_0$ of the $\delta$-function 
barrier that models non-unitary transmission of a point contact (solid curve). The current
unit is $i_0 = e \dot{V}_z / V_{\text{R}}$, and parameters are $\bar\gamma=0.37$ (the
value for GaAs holes), $V_z = 0.1\, V_{\text{R}}$, $E_{\text{F}}=1.2\, E_0$, $B_x=0$,
$B_y = 0.02\, B_0$, and $w/d=2$. For comparison, we also show the current (multiplied
by a factor of 200 to be visible in the plot) generated in a GaAs \textit{electron\/} system
for the same $V_z$, $E_{\text{F}}$, and $\vek{B}$, with band nonparabolicity included
(dot-dashed curve) and neglected (dashed curve).  Currents are scaled by the factor
$d/L$.}
\end{figure}

Figures~\ref{fig:obsCurrent} and \ref{fig:IV} show the dependence of the spin-orbit
gauge-field generated current $i_{\text{LR}}$ on a number of experimentally relevant
quantities. We use $i_0 = e \dot{V}_z / V_{\text{R}}$ as the unit of current and scale
current values by $d/L$  to absorb the linear dependence on $L$.

The magnitude of $i_{\text{LR}}$ as a function of the aspect ratio for the quasi-1D
confinement is shown in Fig.~\ref{fig:obsCurrent}. While there is a precipitous drop
in current as $d/w$ deviates from one, it still remains finite in situations where
the nature of the propagating states becomes more 2D-like.

Figure~\ref{fig:IV} elucidates the dependence of $i_{\text{LR}}$ on the $\delta$-barrier
strength and, thus, on the transmission of the quasi-1D system. For comparison, we
show our result for a hole system alongside two curves that are obtained for
conduction electrons in the same material under the same conditions, with and without
band nonparabolicity included. Two major differences between the conduction-electron
and hole cases are apparent. Firstly, in the same semiconductor material, the maximum
of $i_{\text{LR}}$ for a \textit{p}-type structure is two orders of magnitude larger than that
for an \textit{n}-type one. Secondly, the dependence on the barrier potential is qualitatively
different in the two cases. For holes, the maximum $i_{\text{LR}}$ occurs in the limit
$V_0\to 0$ when the quasi-1D system is perfectly transmitting. In a point contact, this
situation can be realized over a wide range of system parameters at the lowest conductance
plateau. In contrast, the maximum transport current in conduction-electron systems appears
for a non-unitary transmission value, and reaching it in a point contact requires fine-tuning of
parameters in the narrow conductance-step region.

The strong enhancement and different qualitative behavior of $i_{\text{LR}}$ for holes as
compared with electrons can be traced back to the strong mixing between HH and LH
states in a quasi-1D system. The resulting nonparabolicity of hole subbands increases
the sensitivity of Fermi wave vectors and, hence, of quantum phases in the transmission
amplitude, to time dependence of SIA spin splitting. Due to their special nature and
comparable energy scales, HH-LH mixing and HH-LH splitting result in qualitatively
different nonparabolicity effects than those present in conduction-electron systems. To
illustrate this point, we show in Fig.~\ref{fig:IV} the transport current that is obtained when
nonparabolicity in the conduction band~\cite{ogg:pps:66,malcher:sms:86} is taken into
account~\cite{*[{We used the conduction-band-structure parameters given in }] [{.}]
cardona:jap:02} (dot-dashed curve) alongside the result for the same system with
nonparabolicity neglected (dashed curve). It is seen that, for conduction electrons,
nonparabolicity also enhances the ac charge current generated by spin-dependent
electromotive forces, but the enhancement is only a fraction of the maximum current
obtained in the parabolic approximation, and the general shape of the $V_0$-dependence
is similar in the cases with and without nonparabolicity. In particular, the current maximum
still occurs at approximately the same finite value of barrier strength $V_0$. This moderate
influence of nonparabolicity in the conduction-electron case has to be contrasted with the
drastically different shape of the curve obtained for a hole system.

In conclusion, charge currents flowing in quasi-1D hole systems in response to a
time-varying SIA spin splitting are predicted to have a large magnitude (in GaAs,
200-times larger than for conduction electrons) due to strong spin-orbit effects in the
valence band and because of subband nonparabolicity arising from HH-LH mixing.
Proportionality to system size indicates the purely quantum origin of these currents,
and their odd symmetry with respect to transverse in-plane magnetic fields enables
unambiguous extraction of this special current component in experiments.

%

\end{document}